\documentclass[twocolumn,showpacs,preprintnumbers,amsmath,amssymb]{revtex4}

\usepackage{epsfig}
\usepackage{dcolumn}
\usepackage{bm}

\begin{document}
\title{Linear response to perturbation of non-exponential renewal processes}

\author{Francesco Barbi$^1$ }
\author{Mauro Bologna $^2$}
\author{Paolo Grigolini$^{1,3,4}$}
\affiliation{$^1$Dipartimento di Fisica "E.Fermi" - Universit\'{a}
di Pisa, Largo
   Pontecorvo, 3 56127 PISA}

   \affiliation{$^2$Departamento de F\'{i}sica,
Universidad de Tarapac\'{a}, Campus Vel\'{a}squez, Vel\'{a}squez
1775, Casilla 7-D, Arica, Chile} \affiliation{$^3$Center for
Nonlinear Science, University of North Texas, P.O. Box 311427,
Denton, Texas 76203-1427, USA} \affiliation{$^4$Istituto dei
Processi Chimico Fisici del CNR, Area della Ricerca di Pisa, Via
G. Moruzzi, 56124, Pisa, Italy}
\date{\today}
\begin{abstract}

We study the linear response of a two-state stochastic process,
obeying renewal condition, by means of a stochastic rate equation
equivalent to a master equation with infinite memory.  We show
that the condition of perennial aging makes the response to
coherent perturbation vanish in the long-time limit.
   \end{abstract}


\maketitle

Many complex processes generate erratic jumps back and forth from
a state ``on" to a state ``off". We limit ourselves to quoting
ionic channel fluctuations \cite{nadler,hanggiionic,fulinski},
currently triggering the search for a form of stochastic resonance
valid also in the non-exponential case \cite{gocciuk1}, and  the
intermittency of blinking nanocrystals \cite{blinking1,
blinking2}: It has been assessed that the intermittent
fluorescence of these materials obey renewal theory
\cite{blinking3}, namely, a jump from one to the other state, has
the effect of resetting to zero the system's memory. The
non-exponential nature of the distribution of sojourn times makes
this renewal process non ergodic and generates aging effects that
are the object of an increasing theoretical interest
\cite{blinking4,lutz}. Similar properties are found with
surface-enhanced Raman spectra of single molecules
\cite{cannistraro}.

The authors of \cite{europhys} proved that aging, in conflict with
the pioneering work of Ref. \cite{kenkre} claiming the equivalence
between Generalized Master Equation (GME) and Continuous Time
Random Walk (CTRW) \cite{montroll}, makes the GME response to
perturbation different from the CTRW response. These authors,
however, did not propose a new approach to the linear response.

Here we present an approach to the linear response valid in the
aging as well as in the stationary case, as in earlier work
\cite{barkai,gocciuk1}. The key ingredient of the method is to
express the CTRW prescription through a Stochastic Master Equation
(SME), which yields the linear response  under the natural
assumption that the time distribution of collisions is not
significantly affected by perturbation.

Let us consider a two-state renewal process, and, for simplicity,
let us assume that the distribution of times of sojourn in the
state ``on" is the same as the distribution in the state ``off".
We  assign to the Survival Probability (SP) of this process,
$\Psi(t)$, the inverse power law form
\begin{equation}
\label{experimental} \Psi(t) = \left(\frac{T}{T +
t}\right)^{\mu-1},
\end{equation}
with $\mu > 1$. This corresponds to the joint action of the time
dependent rate \cite{gocciuk3,cox} $r(t) = r_{0}/(1 + r_{1} t)$,
with $r_{0} = (\mu-1)/T $ and $r_{1} = 1/T$, and of a resetting
prescription. To illustrate this condition, let us imagine the
random drawing of a number from the interval $I = [0,1]$ at
discrete times $i= 0,1,2...$. The interval $I$ is divided into two
parts, $I_{1}$ and $I_{2}$, with $I_{1}$ ranging from $0$ to
$p_{i}$, and  $I_{2}$ from $p_{i}$ to $1$. Note that $p_{i} = 1 -
q_{i} < 1$ and $q_{i}<<1$, and, as a consequence, the number of
times we keep drawing numbers from $I_{1}$, without moving to
$I_{2}$, is very large.

Let us evaluate the distribution of these persistence times, and
let us discuss under which conditions we get the SP of Eq.
(\ref{experimental}). The SP function is the probability of
remaining  in $I_{1}$ after $n$ drawings, and is consequently
given by
\begin{equation}
\label{towardscox0} \Psi(n) = \prod_{i=1}^{n} p_{i}.
\end{equation}
Using the condition $q_{i} << 1$, and evaluating the logarithm of
both terms of Eq. (\ref{towardscox0}), we obtain
\begin{equation}
\label{towardscox} log (\Psi(n)) = - \sum_{i=1}^{n}  q_{i}.
\end{equation}
The condition $q_{i} << 1$ implies that $i$ and $n$ of Eq.
(\ref{towardscox}) are so large as to make $q_{i}$ virtually
identical to a function of the continuous time $t$,  $q_{i}  =
q(t) = r_{0} \eta(t) $, with $\eta(t) = 1/(1 + r_{1} t)$. Thus,
Eq. (\ref{towardscox}) yields the SP of Eq. (\ref{experimental}),
and the corresponding waiting time distribution density,
$\psi(\tau)$, reads
\begin{equation}
\label{theoretical} \psi(\tau) = (\mu - 1) \frac{T^{\mu -1}}{(\tau
+ T)^{\mu}}.
\end{equation}

We denote as \emph{collisions} the rare drawings of a number from
$I_{2}$, followed by resetting. Thus the collisions occurring at
times $\tau_{1}$, $\tau_{1} + \tau_{2}$, ..., yield: $\eta(t) =
1/(1 + r_{1} t), 0<t<\tau_{1}$; $\eta(t) = 1/(1 + r_{1}
(t-\tau_{1})), \tau_{1} <t<\tau_{1} + \tau_{2}$, and so on. Note
that $\eta(0) = 1$ means that we \emph{prepare} the system at time
$t = 0$.  We adopt a coin tossing prescription to decide whether
to keep or to change sign, after any collision.

The state of the system after the $n$-th collision is described by
the two-dimensional vector ${\bf P}(n)$,  whose components,
$P_{1}$ and $P_{2}$,  are the probabilities of finding the system
in the corresponding states. The $n$-th collision produces the
change
\begin{eqnarray}
\label{natural} {\bf P} (n)  = {\bf M} {\bf P} (n-1),
\end{eqnarray}
where
\begin{eqnarray}
  {\bf M} = \left(\begin{array} {cc}1/2 & 1/2 \\
                          1/2 & 1/2 \end{array}
                          \right) . \end{eqnarray}

Let us denote  by $\sigma_{1}(t)$ and $\sigma_{2}(t)$, the
probabilities  of finding the system in the state ``on" and
``off", respectively, at a time $t$, which, does not necessarily
correspond to the collision occurrence. These probabilities  are
driven by the following SME
\begin{equation}
\label{firstequation2}
\begin{array}{cc}
\frac{d}{dt} \sigma_{1}(t) = - \frac{r_{+}(t)}{2} \sigma_{1}(t) +
\frac{r_{-}(t)}{2} \sigma_{2}(t)\\
\linebreak\\
\frac{d}{dt} \sigma_{2}(t) = - \frac{r_{-}(t)}{2} \sigma_{2}(t) +
\frac{r_{+}(t)}{2} \sigma_{1}(t),\\
\end{array}
\end{equation}
with $r_{+}(t) = r_{-}(t) =  r(t) = r_{0} \xi(t)$. The condition
$r_{+}(t) \neq r_{-}(t)$ is hereby used to discuss the    response
to perturbation. The variable $\xi(t)$ always vanish but in the
correspondence of a collision, where it gets  the value $1/r_{0}$,
so as to make Eq. (\ref{firstequation2}) equivalent to  Eq.
(\ref{natural}),  Note that time $t$ is discrete and that the
derivatives $d\sigma_{i}/dt$, with $i=1,2$, are defined by
$d\sigma_{i}/dt  \equiv \sigma_{i}(t+1) -  \sigma_{i}(t)$. In
conclusion, the collision time is a stochastic variable making the
rate $r(t)$  fluctuate  between $0$ and $1$.

To derive the linear response to perturbation, it is necessary to
connect the SME of Eq. (\ref{firstequation2}) to the CTRW
functions $\Psi(t)$ and $\psi(t)$, including the corresponding
aging properties as well. To establish this connection, let us
notice that the CTRW picture  can be converted into the non-Markov
GME \cite{kenkre,gerardo1}
\begin{equation}
\label{gme} \frac{d}{d t} {\bf P}(t) = - \int_{0}^{t} d \tau
\Phi(t-\tau) {\bf K} {\bf Pp}(\tau).
\end{equation}
The matrix $\bf K$ is defined by
\begin{equation}
\label{matrix} \mathbf{K} \equiv \frac{1}{2}\left(
\begin{array}{rr}
1 & -1 \\
-1 & 1
\end{array}
\right)
\end{equation}
and the memory kernel $\Phi(t)$ is related to $\psi(t)$ in the
Laplace domain through
\begin{equation}
\label{exp} \hat \Phi(u) = \frac{u \hat \psi(u)}{1 - \hat
\psi(u)},
\end{equation}
where $\hat \Phi(u)$ and $\hat \psi(u)$ denote the Laplace
transforms of $\Phi(t)$ and $\psi(t)$, respectively.

Now, let us define the quantity
\begin{equation}
\label{fromheretothere} \Pi(t) = {P_{1}(t) - P_{2}(t)} = 2
P_{1}(t) -1.
\end{equation}
Using Eq. (\ref{gme}), we get for the Laplace transform of
$\Pi(t)$, $\hat \Pi(u)$, the following expression:
\begin{equation}
\hat \Pi(u) = \frac{1}{u +  \hat \Phi(u)},
\end{equation}
which, in turn, using Eq. (\ref{exp}), gives $\hat \Pi(u) = \hat
\Psi(u)$, namely, the  key property
\begin{equation}
\label{basicremark} \Pi(t) = \Psi(t).
\end{equation}

Let us go back to Eq. (\ref{firstequation2}) with no perturbation.
The average on the fluctuating rate yields $<\sigma_{i}(t)> =
P_{i}(t)$, with $i =1,2$, and the average of the stochastic
quantity $\Sigma(t) \equiv \sigma_{1}(t) - \sigma_{2}(t) =
2\sigma_{1}(t) - 1$  becomes equal to $\Pi(t)$ of Eq.
(\ref{fromheretothere}). To double check this important
prediction, let us assume that all the systems are located at the
beginning of their sojourn in either the state ``on" or the state
``off", so as to fit the prescription of preparing the system at
$t= 0$. Thanks to Eq. (\ref{firstequation2}), we obtain

\begin{equation}
\label{sigma} \Sigma(t) = exp\left (-r_{0} \int_{0}^{t}
\xi(t^{\prime}) dt^{\prime}\right ),
\end{equation}
which yields

\begin{equation}
\label{average} \left <  \Sigma(t) \right > = \Psi(t).
\end{equation}
In fact, $\Sigma(t)$ changes only at the occurrence of a
collision, and $\Psi(t)$, as we have seen, is the probability that
no collision occurs up to time $t$. Using Eq. (\ref{basicremark}),
we conclude that $<\Sigma(t)> = \Pi(t)$.

With the help of Eq. (\ref{firstequation2}) we establish the
condition for the linear response to occur. Let us assume
$r_{\pm}(t) = r_{0} \xi(t)(1 + \epsilon F_{\pm}(t))$, where
$\epsilon$ is the perturbation strength and the functions
$F_{\pm}(t) $ describe the action of external perturbation on the
corresponding states. Using Eq. (\ref{firstequation2}) we obtain
\begin{equation}
\label{ratherthan} \frac{d}{dt} \Sigma(t) = - r_{0} \xi(t)(1 +
\epsilon S(t)) \Sigma(t) -r_{0} \xi(t) \epsilon f(t)) ,
\end{equation}
where $S(t) \equiv (F_{+}(t) +F_{-}(t))/2$ and $f(t) \equiv
(F_{+}(t) - F_{-}(t))/2$.  The requirement that the system
response does not vanish enforces the condition $f(t) \neq 0$, and
the additional request that this response be linear yields
$\epsilon S(t)<<1$. Note that the earlier illustrated ideal
numerical experiment, in the presence of perturbation, should be
done with $q_{\pm}(t) = r_{0} \eta(t) (1 + \epsilon F_{\pm}(t))$.
With $q_{\pm}(t)$ remaining very small, the occurrence of
long-time collisions is determined by very small values of
$\eta(t)$, with a very small dependence on $F_{\pm}(t)$, which is
sufficient, however, to produce a time-dependent bias on $\Pi(t)$.
This makes it natural to assume that $\xi(t)$ is independent of
perturbation.

  Using the linear response approximation and solving
the corresponding approximated form of Eq.(\ref{ratherthan}), we
arrive at
\begin{eqnarray}
\Sigma(t) = - r_{0} \epsilon \int_{0}^{t} dt^{\prime} exp
\left(-r_{0} \int_{t^{\prime}}^{t} \xi(t^{\prime \prime})
dt^{\prime
\prime}\right) \xi(t^{\prime}) f ( t^{\prime}) \nonumber\\
+ \Sigma(0) exp \left (- r_{0} \int_{0}^{t} \xi(t^{\prime})
dt^{\prime}) \right ).\hspace{1.8cm}
\end{eqnarray}
We are interested in the mean value of $\Sigma(t)$, namely
$\Pi(t)$. This leads us to write
\begin{equation}
\label{top} \Pi(t) = - \epsilon \int_{0}^{t} \chi(t,t^{\prime}) f
(t^{\prime}) dt^{\prime} + \Pi(0) \Psi(t),
\end{equation}
where
\begin{equation}
\label{derivative} \chi(t,t^{\prime}) = \frac{d}{d t^{\prime}}
\Psi(t,t^{\prime}),
\end{equation}
with $\Psi(t,t^{\prime})$ being the characteristic function
\begin{equation}
\Psi(t,t^{\prime}) \equiv \left <exp (- r_{0} \int_{t^{\prime}}
^{t} dt^{\prime \prime} \xi(t^{\prime \prime}))\right >.
\end{equation}
With $\Pi(0) = 0$, Eq. (\ref{top}) becomes formally identical to
the common linear response prescription \cite{thomas}, with aging
violating, however, the stationary condition $\chi(t,t^{\prime}) =
\chi(t-t^{\prime})$.

Using the same arguments as those earlier adopted to prove Eq.
(\ref{average}), we show that $\Psi(t,t^{\prime})$ coincides with
the survival probability of age $t^{\prime}$ of Ref. \cite{last}.
According to \cite{last}, the function $\Psi(t,t^{\prime})$
coincides with the aging correlation function of the dichotomous
signal under study, $C(t,t^{\prime})$. By means of the stationary
assumption we recover the results of Refs. \cite{barkai,gocciuk1},
and we realize why these earlier results go beyond the Green-Kubo
prediction \cite{fleurov,note}.

In the case $\mu < 2$, the stationary condition cannot be
realized, not even in principle, while in the case $\mu > 2$ it is
possible, even if $\mu < 3$ makes very slow the relaxation to
equilibrium \cite{jacopo}.  Let us study first the more
traditional condition $\mu > 2$, and let us switch perturbation on
after the transient process necessary to reach equilibrium. We
have \cite{aging}
\begin{equation}
\chi(t,t^{\prime}) = - \psi_{\infty}(t- t^{\prime}) = - (\mu - 2)
\frac{T^{\mu-2}}{(t-t^{\prime}+T)^{\mu-1}},
\end{equation}
the stationary correlation function of the ``on" and ``off"
fluctuation, $C(t-t^{\prime})$, exists, with the analytical form
$(T/(t+T))^{\mu-2}$, thereby yielding
\begin{equation}
 \label{stationary}
\psi_{\infty}(t - t^{\prime}) =   -\frac{d}{dt} C(t-t^{\prime}).
\end{equation}
Note that the linear response of Eq. (\ref{top}), with $-
\chi(t,t^{\prime})$ replaced by $\psi_{\infty}(t-t^{\prime})$ of
Eq. (\ref{stationary}) coincides with the prescription of the
phenomenological approach of Ref. \cite{gocciuk1}. In the case
where $f(t) = \Theta(t)$, with $\Theta(t)$ denoting the unit step
function,  by plugging Eq. (\ref{stationary}) into Eq.
(\ref{top}), we obtain
\begin{equation}
\Pi(\infty) = \epsilon,
\end{equation}
and consequently the constant drift of Ref. \cite{barkai}.

In the case where $f(t) = \Theta(t) \cos(\omega t)$, namely the
case of stochastic resonance discussed by the authors of Ref.
\cite{gocciuk1}, by using the expression for the Laplace transform
of $\psi_{\infty}(t)$ and the convolution theorem, we obtain for
the Laplace transform of $\Pi(t)$ the following result:
\begin{equation}
\label{fromklafter} \hat \Pi(u) = \epsilon  (1 + c T^{\mu-2}
u^{\mu-2})\frac{u}{(u^{2} + \omega^{2})},
\end{equation}
where $c \equiv -\Gamma(3-\mu)$. Using the fractional derivative
method of Ref. \cite{bolognone}, we evaluate the anti-Laplace
transform of $\hat \Pi(u)$ of Eq. (\ref{fromklafter}). In the case
$\omega T << 1$, we obtain \cite{details,Tgreat}:
\begin{eqnarray}
\label{asym2} \Pi(t)& \approx &\epsilon \left[1+ \left(\omega T
\right) ^{\mu-2}\Gamma \left(3-\mu \right)\cos \frac{\pi
}{2}\mu\right]
  \cos\omega t \nonumber \\
&   & - \epsilon\left(\omega T \right) ^{\mu-2}\Gamma \left(3-\mu
\right) \sin \frac{\pi }{2}\mu\sin\omega t .
\end{eqnarray}

As earlier pointed out, the theory of this Letter applies also to
the aging case, and to the condition $\mu < 2$, which makes aging
a perennial condition of renewal systems. It is known
\cite{godreche} (see also Ref. \cite{last}) that the exact
expression for the aged $\psi(t,t^{\prime})$
is
\begin{equation}
\psi(t,t^{\prime}) = \psi(t) + \sum_{n=1}^{\infty}
\int_{0}^{t^{\prime}} \psi_{n}(\tau) \psi(t-\tau) d\tau,
\end{equation}
where $\psi_{n}(t)$ denotes the probability density of occurrence,
at time $t$, of the last of a sequence of $n$ collisions. Let us
go back to Eq. (\ref{top}) and let us set $\Pi(0) = 0$. This has
the effect of producing a condition where a non-vanishing $\Pi(t)$
appears only as a perturbation effect. For the Laplace transform
of $\Pi(t)$, we obtain the following expression
\begin{equation}
\label{quote} \hat \Pi(u) = - \epsilon Re (\hat E(u)),
\end{equation}
where $\hat E(u)$ is the Laplace transform of
\begin{equation}
E(t) \equiv \int_{0}^{t} dt^{\prime} \psi(t, t^{\prime}) exp(- i
\omega t^{\prime}).
\end{equation}
After some algebra, we find

\begin{equation}
\label{nostepfunction} \hat E(u) = \frac{i}{\omega}
\frac{\left(\hat \psi(u+i\omega) - \hat \psi(u)\right)}{\left(1 -
\hat \psi(u+i\omega)\right )}.
\end{equation}
It is of some interest  to study Eq. (\ref{nostepfunction}), with
$\mu > 2$ and $\omega T <<1$. Using the imaginary Laplace
transforms of  \cite{imaginary}, for $u \rightarrow 0$ we get an
analytical form for $\hat E(u)$, and with this analytical form
plugged into Eq. (\ref{quote}), we recover Eq.(\ref{asym2}) in the
same limiting condition, namely the Laplace transform of $\epsilon
\cos(\omega t)$. This means that perturbing the system in the
infinitely aged condition, or keeping it under the perturbation
influence, from the preparation to the infinitely aged condition,
yields the same result. It is also interesting to remark that Eq.
(\ref{nostepfunction}) establishes the effect of the step function
perturbation in the case $\mu < 2$. We must use Eq.
(\ref{nostepfunction}) in the limiting case $\omega \rightarrow
0$. This limit condition  yields $\hat E(u) = - (d\hat
\psi(u)/du)/(1 - \hat \psi(u))$, and with it, through the limit
for $u \rightarrow 0$ of $\hat \psi(u)$, with $\mu < 2$, and by
means of Eq.(\ref{quote}), the final result $\hat \Pi(u) =
\epsilon/u$: The asymptotic limit is independent of the value of
$\mu$, and consequently of aging.

The perturbation $f(t) = \Theta(t) \cos(\omega t)$, on the
contrary, produces a response sensitive to the value of $\mu$
adopted. To show this important aspect let us make the assumption
$\omega T <<1$, which,  in the case $\mu > 2$, turns $\hat \Pi(u)$
into the Laplace transform of $\epsilon \cos (\omega t)$, as shown
by Eq. (\ref{asym2}).  In the case $\mu < 2$, the assumption
$\omega T <<1$ allows us to obtain
\begin{equation}
\hat \Pi(u) =  \epsilon Re \left[\frac{i}{\omega} \left(1 -
\left(\frac{u}{u+i\omega}\right)^{\mu-1}\right) \right].
\end{equation}
Using the fractional derivative method of Ref. \cite{bolognone} we
find \cite{details}
\begin{eqnarray}
\Pi(t) &= & -\epsilon Re\left[ \sum\limits_{n=1 }^{\infty}{\mu-1
\choose n}\frac{\left(-\imath \omega t \right)^{n-1}}{(n-1)!}
e^{-\imath \omega t} \right] \\
&       &=-\epsilon Re\left[  \left(\mu -1 \right)F\left(2-\mu,2,
\imath \omega t \right) e^{-\imath \omega t}\right] \nonumber,
\end{eqnarray}
where $F(\alpha, \beta, z)$ is the confluent hypergeometric
function \cite{abramowitz}. In the special case $\mu = 3/2$, of
interest for the physics of blinking quantum dots
\cite{blinking1,blinking2},  the method of Ref. \cite{bolognone}
yields \cite{details}

\begin{equation}\label{38bess}
\Pi(t) =- \frac{ \epsilon}{2}\left[J_0\left( \frac{ \omega t}{2}
\right)\cos\left( \frac{ \omega t}{2} \right)- J_1\left( \frac{
\omega t}{2} \right)\sin\left( \frac{ \omega t}{2} \right)
\right],
\end{equation} with $J{n}()$ denoting the Bessel function, and, more
in general :
\begin{equation}\label{38asimp}
  \Pi(t)\approx\epsilon\frac{\cos\left(\frac{\pi}{2}\mu-  \omega t
  \right)}{\Gamma(\mu-1)(\omega t)^{2+\mu}},
\end{equation}
which means that the condition $\mu < 2$ annihilates  the response
coherence in the long-time limit.

In conclusion, we derive the linear response of non-exponential
renewal processes, without limitation to the stationary condition,
and we show that aging, an unavoidable consequence of $\mu < 2$,
has the interesting effect of annihilating the response to
coherent perturbations, in the long-time limit. This result is
obtained by means of the weak perturbation of a SME, which is
equivalent to perturbing the corresponding CTRW. Along these lines
it would be possible to consider also the case when the collision
occurrence is significantly affected by perturbation. This would
imply, however, a departure from the linear response.

\emph{Acknowledgment} P.G. acknowledges Welch for financial
support through Grant 70525.


\begin{thebibliography}{100}
\bibitem{note}
We note that to recover the Green-Kubo prediction,
$\chi(t,t^{\prime}) = \chi(t-t^{\prime})  \propto
C(t-t^{\prime})$, we should have made the assumption $r_{\pm}(t) =
r_{0} \xi(t) \pm r_{0}\epsilon f(t)$, which does not properly
reflect the physics of these processes.
\bibitem{details} Details on these calculations will be given in a
more extended publication.
\bibitem{Tgreat}  With $\omega T >> 1 $
we obtain $\Pi(t) \approx \epsilon (\mu -2) sin (\omega t)/(\omega
T)$, in qualitative agreement with the resonant-like effect of
Ref. \cite{gocciuk1}, which is expected become quantitative when
the condition $\epsilon \propto log(1/T)$ is taken into account.
\end{thebibliography}
\end{document}